\documentclass[conference]{IEEEtran}
\IEEEoverridecommandlockouts
\usepackage{cite}
\usepackage{amsmath,amssymb,amsfonts}
\usepackage{algorithmic}
\usepackage{graphicx}
\usepackage{textcomp}
\usepackage{xcolor}
\usepackage{ulem}
\usepackage{graphicx}
\usepackage{amsmath}
\usepackage{amssymb}
\usepackage{booktabs}
\usepackage{enumitem}
\usepackage{tabulary}
\usepackage{adjustbox}
\usepackage{silence}
\usepackage{rotating}
\usepackage{lipsum}

\usepackage{caption}
\usepackage{array}
\usepackage{geometry}
\newcommand\blfootnote[1]{%
\begingroup
\renewcommand\thefootnote{}\footnote{#1}%
\addtocounter{footnote}{-1}%
\endgroup
}

\makeatletter
\newcommand{\linebreakand}{%
    \end{@IEEEauthorhalign}
    \hfill\mbox{}\par\mbox{}
    \hfill\begin{@IEEEauthorhalign}
}
\makeatother

\def\BibTeX{{\rm B\kern-.05em{\sc i\kern-.025em b}\kern-.08em
    T\kern-.1667em\lower.7ex\hbox{E}\kern-.125emX}}
\begin{document}
\newgeometry{top=2.54cm, left=1.91cm, right=1.91cm}
\title{Deep3DSketch+$\backslash$+: High-Fidelity 3D Modeling from Single Free-hand Sketches\\
\thanks{This work is supported by xxx}
}

\author{
\IEEEauthorblockN{Ying Zang*}
\IEEEauthorblockA{\textit{School of Information Engineering} \\
\textit{Huzhou University}\\
Huzhou, China \\
02750@zjhu.edu.cn}
\and
\IEEEauthorblockN{Chaotao Ding*}
\IEEEauthorblockA{\textit{School of Information Engineering} \\
\textit{Huzhou University}\\
Huzhou, China \\
02750@zjhu.edu.cn}
\and
\IEEEauthorblockN{Tianrun Chen*}
\IEEEauthorblockA{\textit{College of Computer Science and Technology} \\
\textit{Zhejiang University}\\
Hangzhou, China \\
tianrun.chen@zju.edu.cn}
\and
\IEEEauthorblockN{Papa Mao}
\IEEEauthorblockA{\textit{Mafu Laboratory} \\
\textit{Moxin Tech}\\
Huzhou, China \\
moxin@wingmail.cn}
\and
\IEEEauthorblockN{Wenjun Hu}
\IEEEauthorblockA{\textit{School of Information Engineering} \\
\textit{Huzhou University}\\
Huzhou, China \\
hoowenjun@foxmail.com}
}

\twocolumn[{%
\renewcommand\twocolumn[1][]{#1}%
\maketitle
\begin{center}
\captionsetup{type=figure}

\includegraphics[width=0.8\textwidth]{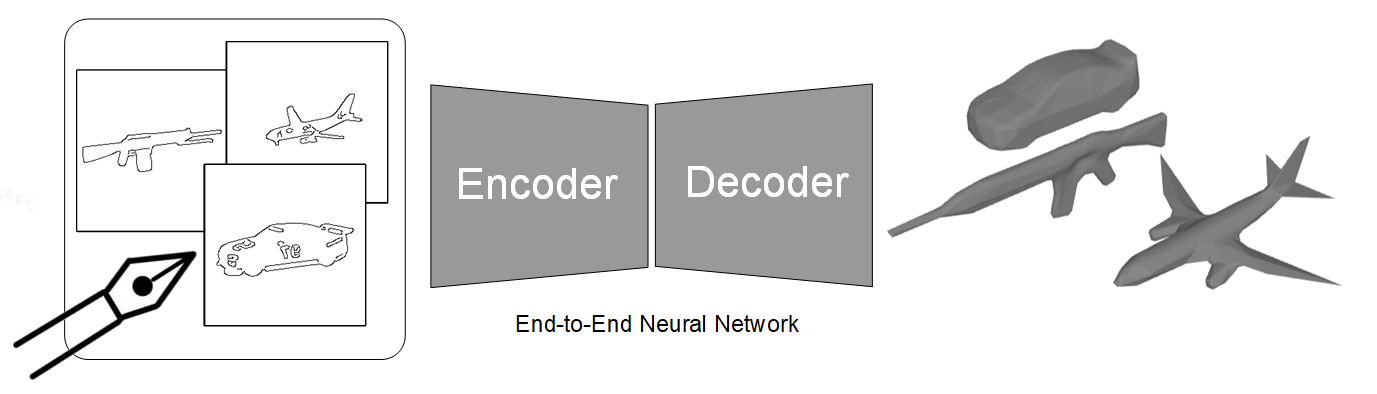}
\captionsetup{font={normalsize}} 
\captionof{figure}{\textbf{The Pipeline of Our Sketch-Based 3D Modeling Approach}. Our approach takes single-view free-hand sketch and feeds it into an end-to-end neural networks, a full 3D model is obtained using the given sketch.}
\vspace{0 in}

\label{fig:teaser}
\end{center}%
}]
\blfootnote{* Equal Contribution}
\blfootnote{$^\top$ Corresponding Author}
\begin{abstract}
The rise of AR/VR has led to an increased demand for 3D content. However, the traditional method of creating 3D content using Computer-Aided Design (CAD) is a labor-intensive and skill-demanding process, making it difficult to use for novice users. Sketch-based 3D modeling provides a promising solution by leveraging the intuitive nature of human-computer interaction. However, generating high-quality content that accurately reflects the creator's ideas can be challenging due to the sparsity and ambiguity of sketches. Furthermore, novice users often find it challenging to create accurate drawings from multiple perspectives or follow step-by-step instructions in existing methods. To address this, we introduce a groundbreaking end-to-end approach in our work, enabling 3D modeling from a single free-hand sketch, Deep3DSketch+$\backslash$+. The issue of sparsity and ambiguity using single sketch is resolved in our approach by leveraging the symmetry prior and structural-aware shape discriminator. We conducted comprehensive experiments on diverse datasets, including both synthetic and real data, to validate the efficacy of our approach and demonstrate its state-of-the-art (SOTA) performance. Users are also more satisfied with results generated by our approach according to our user study. We believe our approach has the potential to revolutionize the process of 3D modeling by offering an intuitive and easy-to-use solution for novice users. 
\end{abstract}

\section{Introduction}
The rapid advancement of AR/VR technology and portable displays has led to an unprecedented demand for 3D content in recent years. Traditionally, 3D content is created by manually implementing Computer-Aided Design (CAD) methods. While CAD software is a powerful tool for creating 3D models, it requires a steep learning curve and a significant investment of time and effort to master. Novice users may struggle with understanding the complex user interface, navigating the software, and manipulating 3D objects. Additionally,
\newgeometry{margin=1.91cm}
even experienced users may face difficulties in creating complex models or dealing with unexpected errors. Despite the challenges, the demand for 3D content continues to grow, and there is a need for more accessible tools and resources to enable novice users to create 3D models easily.

Among the available tools, sketch-based 3D modeling stands out as a promising solution due to its utilization of the intuitive and familiar method of sketching to express ideas. By allowing users to translate their sketches directly into 3D models, the sketch-based 3D modeling approach can significantly reduce the learning curve for new users, making it easier for them to 
create 3D content.

However, existing sketch-based 3D modeling tools are far from perfect. Current methods for sketch-based 3D modeling typically require precise line drawings from various viewpoints or involve a step-by-step workflow that assumes familiarity with breaking down the 3D modeling process into individual steps \cite{cohen1999interface,deng2020interactive}. These methods, while effective, are not novice-friendly and can be time-consuming. Other works employing template primitives or retrieval-based approaches \cite{chen2003visual, wang2015sketch} lack the ability to provide full customizability to users. As a result, there is a need for an approach to 3D modeling that balances ease of use and flexibility, allowing novice users to create custom 3D models with minimal effort and maximum creative freedom.

To fulfill the goal of rapid and intuitive 3D modeling, our work presents a groundbreaking approach that leverages a single sketch as the input to generate a comprehensive and intricately detailed 3D model. This task poses significant challenges, especially due to the constraint of using just a single free-form sketch as the sole input. The sketches used as input are often sparse and ambiguous. The sparsity of sketches arises due to the fact that they offer only a single view, which limits the amount of information available for constructing a complete 3D model. Furthermore, the lack of texture information in sketches makes depth estimation a challenging task, leading to a significant amount of uncertainty when attempting to learn 3D shapes. The ambiguity of sketches is another significant challenge. Since sketches are abstract, the same set of strokes can have different interpretations in the 3D world, leading to multiple possible interpretations of the same sketch. Additionally, sketches are often abstract and lack fine boundary information when drawn manually, making it difficult to discern the precise shape and form of the intended object. As a result, it is important to develop robust methods for handling ambiguity to ensure that the generated 3D models align with the user's intended design.

To address this challenge, we present a pioneering end-to-end 3D modeling network, which effectively tackles the task at hand. \textbf{Deep3DSketch+$\backslash$+}. The overall pipeline for this approach is illustrated in Fig.1. We first introduce symmetric clues to resolve the ambiguity issue. We leverage the prior knowledge that man y real world objects are bilaterally symmetric about a reflection plane \cite{liu2010computational, zhou2021nerd}, so the reflection symmetry provides a strong geometric constraint that can effectively reduce the uncertainty in the 3D modeling process. Therefore, we constrain the vertices of the reconstructed object to be symmetric, and ensure that the mesh rendered images reflect this symmetry. To overcome the issue of sparsity, we propose the incorporation of structural clues through a structural-aware Shape Discriminator (SD). The SD takes input from both the predicted mesh and ground truth shape, enhancing the network's capability to generate realistic 3D models. The SD takes as input rendered silhouettes that are sampled at random views, which provides the best coverage for representing both the generated mesh and ground truth mesh.

By leveraging reflection symmetry in single-view 3D modeling, we not only reduce the complexity of the modeling process but also address the ambiguity issue. We extensively evaluated our approach through a series of experiments, demonstrating its effectiveness in achieving state-of-the-art (SOTA) performance on synthetic and real datasets. Our method exhibits higher fidelity and excels in capturing detailed structural information. We also performed user study that users are more satisfied with 3D models generated by our approach compared to existing methods. We believe that our method brings significant advancement in the field of 3D modeling and has promising applications in industries such as architecture, engineering, and design. 

\begin{figure*}
\centering
\includegraphics[width=1\textwidth]{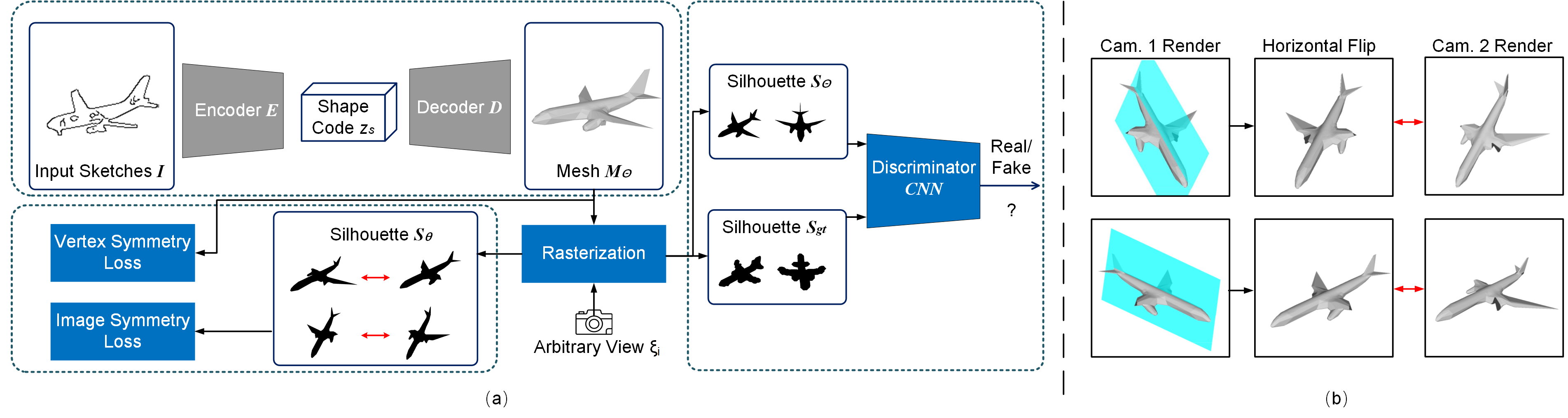}
\caption{(a) \textbf{The Architecture of Deep3DSketch+$\backslash$+.} The backbone encoder-decoder structure takes the input of a sketch and generate a mesh. The mesh is supervised by the rendered silhouette and a shape discriminator. (b) A mesh is rendered with two cameras using differential rendering. The second camera's viewpoint is obtained by reflecting the first camera's viewpoint about the mesh's plane of symmetry (blue). The resulting second render is then compared to a horizontally flipped version of the first render.}
\end{figure*}

\section{Related works}
\subsection{Sketch-based 3D Modeling}
Sketch-based 3D modeling has been an area of active research for several years, with some approaches proposed by researchers. One category of sketch-based 3D modeling approaches is interactive approaches, which involve decomposing the task into sequential steps or requiring specific drawing gestures or annotations. These approaches have been shown to require a significant amount of strategic knowledge, which can be challenging for novice users. Examples of interactive approaches include the work of Li et al. \cite{li2020sketch2cad}, which decomposes the task into two stages of coarse-to-fine reconstruction, and the work of Cohen et al. \cite{cohen1999interface}, which uses annotation-based feedback to refine the 3D model.In contrast, end-to-end approaches, such as those that use template primitives or retrieval-based methods, tend to be more straightforward but lack customizability. These approaches involve generating the 3D model directly from the sketch without intermediate steps. An example of an end-to-end approach is the work of Chen et al. \cite{chen2003visual}, which proposes the use of 3D geometric primitives for sketch-based modeling. Wang et al. \cite{wang2015sketch} introduced a retrieval-based approach that uses a database of 3D models to find the closest match to the input sketch. In recent times, there has been a surge in the development of deep learning-based methods for single-view 3D reconstruction, which also encompasses sketch-based 3D modeling. Zhang et al. \cite{zhang2021sketch2model} and Guillard et al. \cite{guillard2021sketch2mesh} proposed methods that employ deep learning to directly reconstruct the 3D model. Nevertheless, these approaches encounter significant challenges due to the inherent sparsity and abstractness of sketches. Sketches often lack fine boundary information and texture details necessary for accurate depth estimation, thereby making it challenging to generate high-quality 3D shapes. To tackle these limitations, our work introduces additional symmetric clues and a shape discriminator, which contribute to enhancing the quality of the generated 3D models.

\subsection{Single-View 3D Reconstruction}
The task of single-view 3D reconstruction has been a challenging problem in the field of computer vision and computer graphics for a long time. With the advent of large-scale datasets like ShapeNet~\cite{chang2015shapenet}, data-driven approaches have gained popularity in recent years. In the field of data-driven methods, certain works~\cite{chen2019learning, park2019deepsdf, chen2023deep3dsketch} leverage category-level information to infer 3D representations from a single image. Other works~\cite{liu2019soft,liu2019learning,kato2018neural} directly generate 3D models from 2D images, and differentiable rendering techniques have played a significant role in achieving this. More recently, there have been proposals for unsupervised methods for implicit function representations utilizing differentiable rendering techniques~\cite{lin2020sdf, yu2021pixelnerf}. As for shape representation, most of these works use mesh-based representation for 3D shapes. Unlike other representations \cite{zhang2023dyn,zhang2023painting,fu2022panoptic, dou2020top, dou2022tore, dou2022coverage, xu2022rfeps, lin2023patch, wang2022progressively, yang2023neural}, mesh representation can be directly integrated to existing shape editing tools.

However, the majority of existing methods primarily focus on learning 3D shapes from 2D RGB images. In contrast, our goal is to generate 3D shapes from 2D sketches, which are more abstract and sparse compared to images. The challenge lies in achieving the generation of high quality 3D shapes from such a sparse and abstract form of input. Sketches lack important information like texture, lighting, and shading, making it difficult to infer 3D geometry accurately. Moreover, sketches are often incomplete, and the same set of strokes can have different interpretations in 3D, adding ambiguity to the problem. Hence, it is crucial to develop a method that can learn to interpret and reconstruct 3D shapes from sparse and ambiguous sketches accurately. In this work, we propose a novel approach that tackles these challenges and provides an efficient and accurate solution for sketch-based 3D modeling.

\section{Methods}

\subsection{Preliminary}

We employ a single binary sketch ${I\in \left\{0,1\right\}^{W\times H}}$ as the input for 3D modeling. Here, ${I \left [ i,j \right ] = 0 }$ indicates a marked stroke, while ${I\left [ i,j \right ] = 1 }$ represents other areas. Our network $G$ is designed to generate a mesh ${M_\Theta =(V_\Theta, F_\Theta)}$, where ${V_\Theta}$ and ${F_\Theta}$ correspond to the mesh vertices and facets, respectively. The silhouette ${S_\Theta :\mathbb{R}^3 \rightarrow \mathbb{R}^2} $ of ${M_\Theta}$ should align with the input sketch $I$.

\subsection{The Auto-Encoder Backbone}

As illustrated in Fig. 2, our proposed method, Deep3DSketch+$\backslash$+, is comprised of an encoder-decoder network structure, where the encoder $E$ transforms the sparse and the objective is to transform an ambiguous input sketch into a latent shape code $z_s$, which captures the essence of the sketch at a coarse level, taking into account the semantic category and conceptual shape. The decoder $D$ is then used to transfer the latent shape code $z_s$ to the mesh $M_\Theta = D(z_s)$. Instead of using structures like MLP to predict point-wise locations, To obtain the output mesh $M_\Theta$, we employ cascaded upsampling blocks that calculate the vertex offsets of a template mesh and deform it accordingly. 
For network supervision, we employ a multi-scale mIoU (Intersection over Union) loss $\mathcal{L}_{sp}$ , which quantifies the similarity between the rendered silhouettes and ground truth silhouettes. To improve computational efficiency, we employ a progressive strategy to incrementally enhance the resolutions of the silhouettes during the training process. This enables a more accurate representation of the object's boundaries and details. 
\begin{align}
\mathcal{L}_{s p}=\sum_{i=1}^{N} \lambda_{s_{i}} \mathcal{L}_{iou}^{i}
\label{lsp}
\end{align}
where $i$ represents the index used to calculate the mIoU loss values for N scales. $\mathcal{L}_{iou}$ is defined as:
\begin{align}
\mathcal{L}_{i o u}\left(S_{1}, S_{2}\right)=1-\frac{\left\|S_{1} \otimes S_{2}\right\|_{1}}{\left\|S_{1} \oplus S_{2}-S_{1} \otimes S_{2}\right\|_{1}}
\end{align}
where $S_1$ and $S_2$ is the rendered silhouette. 

\subsection{Leveraging the Symmetry Prior}
The silhouette constraint ensures the neural network generates a 3D mesh matching the input sketch from a specific viewpoint. However, as the model exists in 3D space, considering other viewpoints is crucial to avoid ambiguity.

To address this, we leverage extra clues. Specifically, we utilize bilateral symmetry as a powerful geometric constraint, exploiting its common property in real-world objects. This constraint effectively reduces uncertainty in the 3D modeling process \cite{leung2022black, zhou2021nerd}. We enforce vertex symmetry and ensure the rendered mesh images display this symmetry.

Therefore, we first introduce the Vertex Symmetry Loss $L_{Vsym}$, which encourages symmetric
mesh vertices according to
\begin{align}
L_{Vsym}=\frac{1}{N}  {\textstyle \sum_{i=1}^{N}} min_V{_j}\left \| T v_i- v_j \right \|_2^2
\end{align}
where $v_i$ are the mesh vertices and the transformation $T$ is 
\begin{align}
T & = I-2\overrightarrow{n} \overrightarrow{n}^\top 
\end{align}
where 
$\overrightarrow{n} \in \mathbb{R} ^3$ is the unit normal vector of the reflection plane. The  $L_{Vsym}$ penalizes distances between each vertex and its nearest neighbor upon reflection about the symmetric plane and force the network to learn the symmetric result. 

Next, we introduce the Image Symmetry Loss $L_{Isym}$ that promotes the image projections that exhibit object symmetry. In practice, for every viewpoint sampled $\xi_{1...N}$, the mesh is rendered differentiably using two cameras. The viewpoint of the second camera is obtained by reflecting that of the first camera about the mesh's plane of symmetry . The output of the second render is then compared to the horizontally flipped output of the first render. We get $S_{\theta}\left \{\xi_1, \xi_{T1}; \xi_2, \xi_{T2}; ...\xi_N, \xi_{TN} \right \} $. $L_{Isym}$ is defined as
\begin{align}
L_{Isym}=\frac{1}{m}\sum_{i=1}^m \sum_{j,k} \left || \gamma (S_{\theta\xi_i})_{j,k}-\gamma (S_{\theta\xi_{Ti}})_{j,k} \right ||^2_2
\end{align} 
This is similar to comparing a simulated image of how the mesh should appear with symmetry, and the network takes the symmetry into consideration. 

\subsection{Structural-Aware Discriminator}
In the process of utilizing the symmetry prior, we generate silhouettes of the reconstructed mesh from multiple viewpoints. Acquiring multi-view silhouettes plays a crucial role in 3D reconstruction tasks, as it presents a unique challenge compared to 2D image translation due to the need to capture the mesh from different angles. To further enhance our approach, we incorporate the use of multi-view silhouettes to exploit structural information. Drawing from previous research in the field of shape-from-silhouette, we recognize the valuable geometric information embedded in multi-view silhouettes regarding the 3D shape. By incorporating these additional views, we can effectively capture the underlying structure and enhance the accuracy of our 3D reconstruction process~\cite{gadelha2019shape, hu2018structure}. Specifically, 
we propose a Shape Discriminator that considers the input of $S_{\theta}\left \{1...N \right \} $ from the predicted mesh $M_{\theta}$ and the silhouettes $S_{gt}\left \{1...N \right \} $ from the ground truth mesh $M_{gt}$. The shape discriminator is a convolutional neural network that learns the distribution of ground truth mesh and calculate a GAN loss to supervise the generation process. Here, we use the non-saturating GAN loss \cite{mescheder2018training}.
\begin{align}
\begin{split}
\mathcal{L}_{sd} &=\mathbf{E}_{\mathbf{z_v} \sim p_{z_v}, \xi \sim p_{\xi}}\left[f\left(SD\left(R(M, \xi)\right)\right)\right] \\
&+\mathbf{E}_{\mathbf{z_{vr}} \sim p_{z_{vr}}, \xi \sim p_{\xi}}\left[f\left(-SD(R(M_r, \xi))\right)\right] \label{gan}
\end{split}\\ 
& \mathit{{ where  }}  f(u)=-\log (1+\exp (-u))
\end{align}

\subsection{Loss Function}
We carefully design the loss functions with five components to train the network: 1) a multi-scale mIoU loss $\mathcal{L}{sp}$, 2) flatten loss and Laplacian smooth loss $\mathcal{L}{r}$, 3) a structure-aware GAN loss $\mathcal{L}{sd}$, 4) the vertex symmetry loss $\mathcal{L}{Vsym}$, and 5) the image symmetry loss $\mathcal{L}_{Isym}$. Laplacian smooth loss and flatten loss , denoted as $\mathcal{L}_{r}$, are employed to enhance the visual quality and realism of the meshes, as demonstrated in previous works such as~\cite{zhang2021sketch2model,kato2018neural,liu2019soft}.

The overall loss function, denoted as $Loss$, is computed as the weighted sum of the aforementioned five components, providing a mathematical expression as follows:
\begin{align}
Loss =  \mathcal{L}_{sp} + \mathcal{L}_{r} + \lambda_{sd} \mathcal{L}_{sd} + \lambda_{sv} \mathcal{L}_{Vsym}+ \lambda_{si}\mathcal{L}_{Isym}
\label{loss}
\end{align}

\section{Experiment}

\begin{figure*}
\centering
\includegraphics[width=0.95\textwidth]{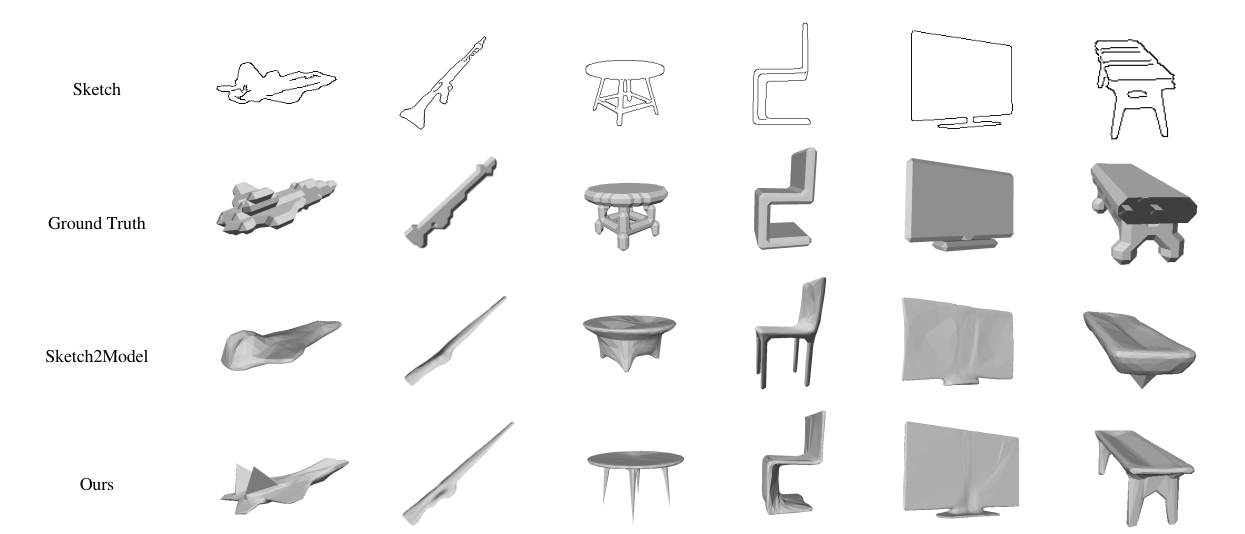}
\caption{\textbf{The comparison of Deep3DSketch+$\backslash$+ with existing approach.} It can be found that Deep3DSketch+$\backslash$+ preserves more details and obtains the 3D model with higher fidelity. }
\vspace{-0.1cm}
\end{figure*}

\subsection{Datasets}
There is a scarcity of publicly available datasets that contain both sketches and their corresponding 3D models. However, in \cite{zhang2021sketch2model}, an alternative approach was taken where synthetic data from the ShapeNet-synthetic dataset was used for training. The synthetic data is generated by employing a canny edge detector on rendered images from Kar et al. \cite{kar2017learning}. 

\subsection{Implementation Details}
We utilized the SoftRas algorithm \cite{liu2019soft} for rendering silhouettes and the ResNet-18 architecture \cite{he2016deep} as the encoder $E$ for extracting image features. For each 3D object in the dataset, we utilized a fixed distance from the camera, an elevation angle of 0 degrees, and an azimuth angle of 0 degrees in the canonical view. The ground-truth viewpoint was used for rendering purposes. To calculate the intersection over union (IoU) loss, we rendered the silhouettes of each predicted and ground truth model using N=4 for rendering. 
The Adam optimizer was used with an initial learning rate of 1e-4, decayed by a factor of 0.3 every 800 epochs, with beta values set to 0.9 and 0.999. We trained the model individually for each class of the dataset, conducting a total of 2000 training epochs. In Equation \ref{loss}, we set $\lambda_{sd}$, $\lambda_{si}$, and $\lambda_{sv}$ to 0.1. The model was trained and evaluated on four NVIDIA GeForce RTX3090 GPUs.

\subsection{Experimental Result and Performance Comparison}
\subsubsection{The ShapeNet-Synthetic Dataset}

We assessed the performance of our approach by comparing it to the model retrieval technique using features from a pre-trained sketch classification network and the current state-of-the-art (SOTA) model, following the identical protocol as described in \cite{zhang2021sketch2model}. Our experiments were conducted on the ShapeNet-Synthetic dataset, which offers precise ground truth 3D models for both training and evaluation purposes. To assess the fidelity of the generated meshes, we used the voxel IoU metric, a commonly used measure for 3D reconstruction, and the results are presented in Table \ref{table:table1}. Our approach was quantitatively evaluated and demonstrated its effectiveness by achieving state-of-the-art (SOTA) performance. Additionally, we compared our method with existing state-of-the-art models, and the results further demonstrated its effectiveness in producing higher-quality models with improved structural fidelity, as illustrated in Fig.3.


\begin{table*}[ht]
\caption{The quantitative evaluation of ShapeNet-Synthetic dataset}

\begin{center}

\scalebox{0.84}{
\begin{tabular}{|c|c|c|c|c|c|c|c|c|c|c|c|c|c|c|}
\hline
\multicolumn{15}{|c|}{Shapenet-Synthetic (Voxel IoU $\uparrow$)} \\
\hline
  & { Cabinet}        & { Bench}          & { Display}        & { Loudspeaker}    & { Telephone}      & { Rifile}         & { Watercraft}     & { Lamp}           & { Airplane}       & { Sofa}           & { Table}          & { Chair}          & { Car}            & { mean}           \\ \hline
  Retrieval    & 0.518    & 0.380        & 0.385     & 0.468        & 0.622         & 0.475          & 0.422      & 0.325       & 0.513        & 0.483         & 0.311             & 0.346     & 0.667         & 0.455          \\ \hline

Auto-Encoder       & 0.663       & 0.467        & 0.541    & 0.629       & 0.706          & 0.605           & 0.556             & 0.431             & 0.576             & 0.613     & 0.512       & 0.496       & 0.769        & 0.582          \\ \hline
{ sketch2model}  & { \textbf{0.701}} & { 0.481}          & { \textbf{0.604}} & { \textbf{0.641}} & { 0.719}          & { 0.612}          & { \textbf{0.586}} & { \textbf{0.472}} & { 0.624}          & { 0.622}          & { 0.478}          & {  \textbf{0.522}}          & { 0.751}          & { 0.601}          \\ \hline

{ Ours} & { \textbf{0.701}} & { \textbf{0.519}} & { 0.593}          & { 0.619}          & {  \textbf{0.755}}          & { \textbf{0.638}} & { 0.576}          & { 0.461}          & { \textbf{0.633}} & { \textbf{0.641}} & { \textbf{0.526}} & { 0.519}          & { \textbf{0.793}} & { \textbf{0.613}} \\ \hline
\end{tabular}}
\end{center}

\label{table:table1}

\end{table*}

%
%

\subsubsection{User Study for 3D Modeling Result} 
Next, we conducted another user study to verify the image quality of our generated images. We conducted a user study following the settings of \cite{cai2021unified,michel2022text2mesh} and used the metric of widely-used Mean Option Score (MOS) ranging from 1-5 \cite{seufert2019fundamental} to the following three factors:
\begin{enumerate}
    \item Q1:  How well does the output 3D model match the input sketch? \textit{(Fidelity)} ; 
    \item Q2:  How do you think the quality of the output 3D model? \textit{(Quality)}.
\end{enumerate}

We recruited 12 designers who were familiar with 3D content and presented 36 generated 3D modeling result by our algorithm to them. Prior to the experiment, we gave each participant a brief and one-to-one introduction to the concept of fidelity and quality. We report the rating result and average the scores. The result in shown in Table \ref{table:tablev}. As perceived by users, our method outperforms existing state-of-the-art method in users' subject ratings.
\begin{table}[h]
\setlength{\tabcolsep}{6mm}
\caption{Mean Opinion Scores (1-5) for Q1 (Fidelity) and Q2 (Quality)}
\begin{center}
\scalebox{1}{
\begin{tabular}{|l|c|c|}
\hline
             & {(Q1):Fidelity} & {(Q2):Quality} \\ \hline
             
Sketch2Model & 3.33                               & 3.12                              \\ \hline
Ours         & \textbf{3.41}                      & \textbf{3.31}                     \\ \hline
\end{tabular}}
\end{center}
\label{table:tablev}
\end{table}

\subsection{Ablation Study}

In order to demonstrate the effectiveness of our proposed method, we performed an ablation study where we removed the Symmetry Prior (SP) using $L_{Isym}$ and $L_{Vsym}$.We also removed the progressive Shape Convolutional Discriminator (SD) for structural awareness. Our quantitative result (Table \ref{table:table3}) shows removing the SP and SD will be detrimental to the performance. The results in Fig.4 further illustrate the superiority of our method over the baseline.

\begin{table}[!htb]
\caption{Ablation Study.}
\begin{center}
\scalebox{0.8}{
\begin{tabular}{c c || c c c c c c c }
\hline
SD & SP & car & sofa & airplane & bench & display & chair & table \\
\hline
\hline
 & & 0.767  & 0.630  & 0.633 & 0.503 & 0.586 & 0.524  & 0.493  \\
$\surd$ &  & 0.782  & 0.640  & 0.632 & 0.510  & 0.588  & \textbf{0.525}  & 0.510  \\

$\surd$ & $\surd$ & \textbf{0.793}  & \textbf{0.641}  & \textbf{0.633} & \textbf{0.519}  & \textbf{0.593}  & 0.519  & \textbf{0.526}  \\

\hline
SD & SP & telephone & cabinet & loudspeaker & watercraft & lamp & rifile & mean \\
\hline
\hline
 & & 0.742 & 0.690 & 0.555 & 0.563 & 0.458 & 0.613 & 0.598 \\
$\surd$ &  & \textbf{0.757}  & 0.699 & \textbf{0.630}  & \textbf{0.583}  & \textbf{0.466} & 0.624  & 0.611  \\

$\surd$ & $\surd$ & 0.755  & \textbf{0.701}  & 0.619 & 0.576  & 0.461  & \textbf{0.638}  & \textbf{0.613}  \\

\hline
\end{tabular}}
\end{center}
\label{table:table3}

\end{table}
\begin{figure}[h]
\centering
\resizebox{.5\textwidth}{!}{
\includegraphics[width=0.7\linewidth]{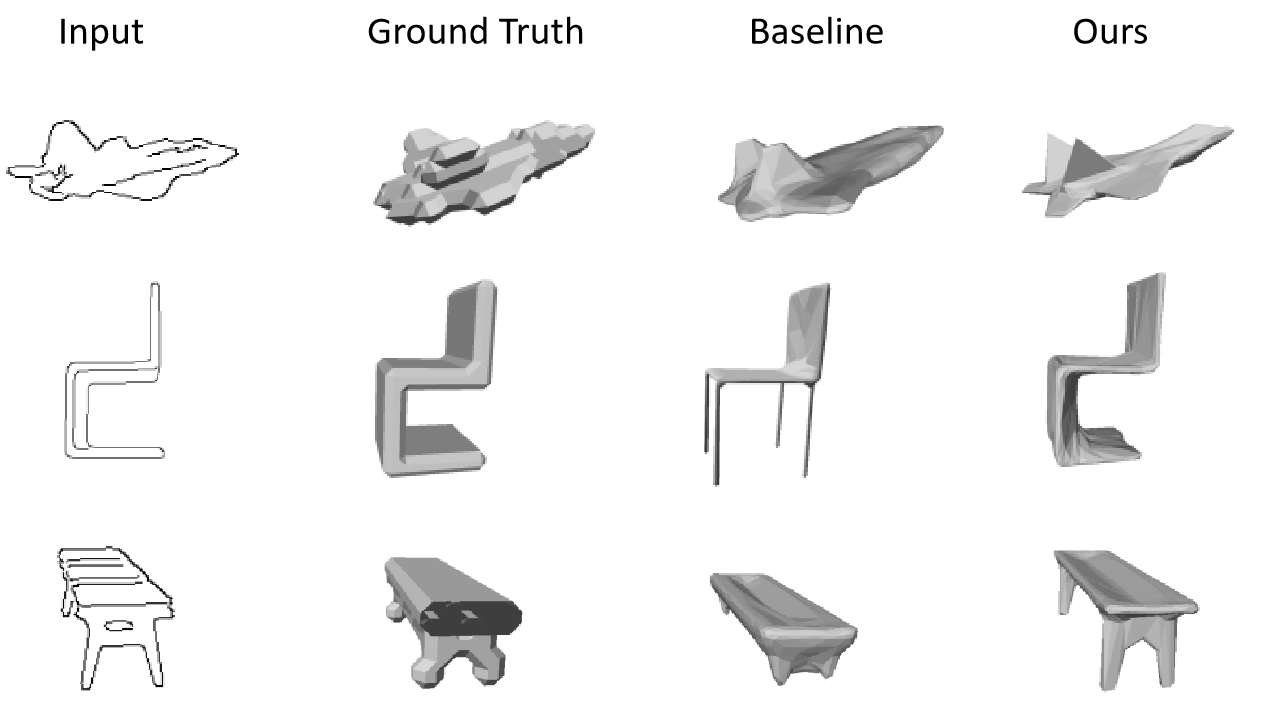}}
\caption{{\textbf{Ablation study.} In contrast to the baseline method, our approach generates more intricate structures and enhances sketch perception ability.}}
\label{fig:4}
\vspace{0.1cm}
\end{figure}

\section{Conclusion}
In this research paper, we introduce Deep3DSketch+$\backslash$+, a novel approach for creating high-fidelity 3D models using a single free-hand sketch as input. Traditional Computer-Aided Design (CAD) methods can be time-consuming and complex, making the creation of 3D models a challenging task. Our method offers a more intuitive and efficient solution by utilizing a end-to-end neural network, which address the issue of sparsity and ambiguity of using single-sketch to perform 3D modeling. Deep3DSketch+$\backslash$+ has leverage the symmetry prior of 3D models to resolve ambiguity issue and the structural-aware shape discriminator that takes symmetric sampling of silhouette of 3D models as the input to resolve the sparsity issue. Our extensive experiments validate the remarkable performance of the proposed Deep3DSketch+$\backslash$+ approach, which outperforms existing methods on both real-world and synthetic data, establishing it as the state-of-the-art (SOTA) solution. We strongly believe that this innovative method has the potential to revolutionize the future of 3D modeling pipelines, enhancing their intuitiveness and accessibility.

\bibliographystyle{IEEEtran}
\bibliography{egbib}

\end{document}